\documentclass[12pt]{iopart}

\usepackage{dcolumn}
\usepackage{bm}

\expandafter\let\csname equation*\endcsname\relax
\expandafter\let\csname endequation*\endcsname\relax

\usepackage{amsmath}
\usepackage{amsfonts}
\usepackage{amssymb}
\usepackage{array}
\usepackage{graphicx}
\usepackage{textcomp}
\newcommand*{\eq}[1]{Eq.~(\ref{#1})}

\begin{document}

\title{Entanglement generation in quantum networks of Bose-Einstein condensates}

\author{Alexey N. Pyrkov$^{1}$ and Tim Byrnes$^{1}$}

\address{$^1$ National Institute of Informatics, 2-1-2
Hitotsubashi, Chiyoda-ku, Tokyo 101-8430, Japan}

\begin{abstract}
Two component (spinor) Bose-Einstein condensates (BECs) are considered as the nodes of an interconnected
quantum network.  Unlike standard single-system qubits, in a BEC the quantum 
information is duplicated in a large number of identical bosonic particles, thus can be considered
to be a ``macroscopic'' qubit.  One of the difficulties with such a system is how to effectively 
interact such qubits together in order to transfer quantum information and create entanglement. 
Here we propose a scheme of cavities containing spinor BECs coupled by optical fiber in 
order to achieve this task. We discuss entanglement generation and 
quantum state transfer between nodes using such macroscopic BEC qubits. 
\end{abstract}

\maketitle

\section{Introduction}

Quantum networks have recently attracted much interest due to their potential applications in quantum computing, communication, metrology, and simulation~\cite{kimble08,duan10,ladd10}.  Quantum information is generated, stored and processed in individual quantum systems (quantum nodes) and connected via quantum channels. One of the attractive 
features of quantum networks is in the flexibility that such a method offers in terms of the way that the quantum 
nodes are connected.  Typically photonic channels are used for transmission of quantum information. Photons are suitable for the quantum channels due to their ability to carry quantum information 
over long distances with modest decoherence.  Dedicated tasks like quantum key distribution can already be achieved using send-only emitter nodes and recieve-only detector nodes~\cite{eisaman11}.  Two well-studied approaches for quantum nodes are atomic ensembles~\cite{duan01} and single atoms~\cite{duan10}. Recently the experimental realization of an elementary quantum network of single atoms in optical cavities was presented~\cite{ritter12}. 

Meanwhile, advances in atom chip technology have led to the ability of precise coherent control of two component Bose-Einstein condensates~\cite{bohi09,riedel10}. In these works, a combination of microwave and radio frequency
pulses were used to control the hyperfine states of $ \sim 10^3 $ atoms coherently, such that an arbitrary position on the 
collective Bloch sphere could be realized.  Furthermore, strong coupling between the excited state of a $^{87}\mbox{Rb}$ BEC~\cite{steck08} and an optical cavity has been achieved~\cite{colombe07}, as well as generation of remote entanglement between a single atom inside an optical cavity and a BEC~\cite{lettner11}. 

On the theoretical side, we have previously studied the possibility of performing quantum computation based on such ``BEC qubits'' \cite{byrnes12}. The primary difference between such two-component BECs and other proposals for
quantum information processing is in the nature of how qubit information is encoded physically.  This can be
seen most simply for a single BEC qubit, where we encode the standard qubit state $\alpha|0\rangle+\beta|1\rangle$ in the following state 
\begin{equation}
\label{becqubit}
|\alpha,\beta\rangle\rangle\equiv\frac{1}{\sqrt{N!}}(\alpha a^\dagger+\beta b^\dagger)^{N}|0\rangle,
\end{equation}
where creation operators for the two hyperfine states $ a^\dagger, b^\dagger $ obey bosonic commutation relations, 
and $ N $ is the number of bosons in the BEC. The state (\ref{becqubit}) is a spin coherent state and has been studied 
in several contexts before \cite{gross12}. The above spin coherent state lives in a $ N + 1 $ dimensional
Hilbert space and is therefore clearly not a genuine two-level qubit system. Nevertheless, it has many 
analogous properties to standard qubits, which were exploited in the proposal of Ref. \cite{byrnes12}. 
 Encoding quantum information in the form (\ref{becqubit}) is attractive
due to the great redundancy that the state possesses.  For example, in a typical BEC $ N = 10^3$-$10^6$ and thus
the same qubit information is duplicated a very large number of times.  Therefore, even in the presence of losses
and decoherence, the quantum information is not easily lost due to the sheer number of copies of the state.  
The problem is then how the state (\ref{becqubit}) can be used perform quantum information tasks, since 
it is not strictly equivalent to a qubit state. A priori it is not clear how, or if it is possible at all, to perform tasks such 
as quantum algorithms using such BEC qubits.  In Ref. \cite{byrnes12} it was shown that such states could
be manipulated analogously to standard qubits, and quantum algorithms could be implemented when many 
such BEC qubits are combined together. One of the key results of Ref. \cite{byrnes12} is that, in analogy with standard
qubits, one and two BEC qubit interactions are necessary in order to perform universal quantum computation. It was also shown 
that tasks such as quantum teleportation is possible using BEC qubits \cite{pyrkov13}. All of these protocols rely on the 
presence of a two qubit gate, most usually of the form $ S^z_i S^z_j $.  Currently, no schemes for implementing such an interaction 
exist, although works proposed originally for single atomic systems, such as Ref. \cite{treutlein06b}, could be generalized to the BEC case. 

In this paper
we show how to create a $ S^z_i S^z_j $ entangling interaction between the nodes of the quantum network, which is fundamental 
to many tasks quantum information processing tasks described above. 
We analyze the performance of the scheme under imperfect conditions with decoherence inducing processes such as spontaneous emission, photon loss, and general dephasing. This paper is organized as follows.  
In section \ref{sec:entgen} we introduce the basic protocol for entanglement generation and show that it creates a $ S^z_i S^z_j $ Hamiltonian.  In section \ref{sec:dec} we discuss the effects of decoherence, induced by spontaneous emission and photon loss.  Experimentally achievable parameters are estimated in section \ref{sec:numbers} based on the decoherence estimates.  Finally, we conclude and summarize our findings in section \ref{sec:conc}.

\begin{figure}
\begin{center}
\scalebox{0.4}{\includegraphics{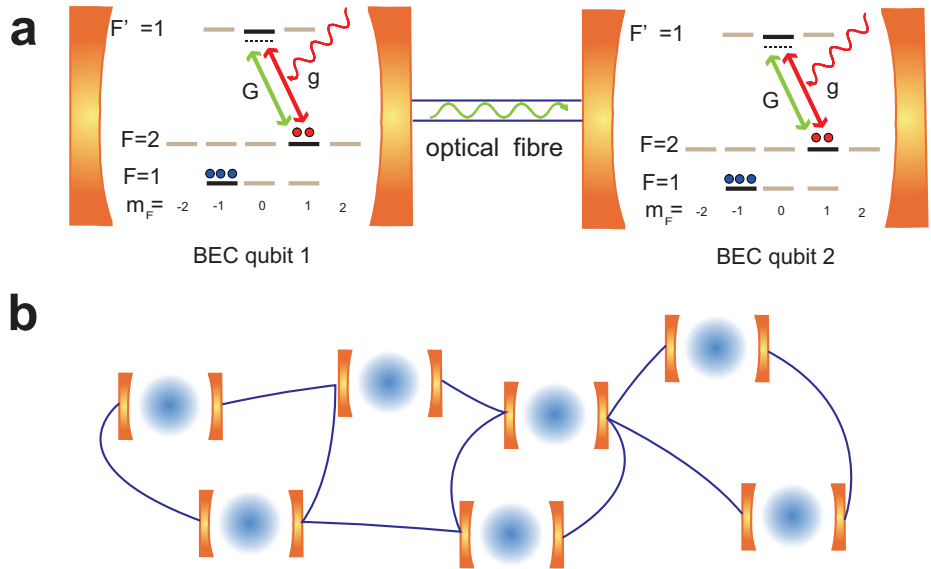}}
\caption{\label{fig1}
(a) Entanglement generation scheme for two BEC qubits. The cavities are off-resonantly tuned to the  $ F'= 1 \leftrightarrow  F= 2 $ transition with coupling $ G $, which allows for photons to propagate between the cavities via the optical fiber.  To generate entanglement between two nodes, an off-resonant laser with transition amplitude $ g $ illuminates the BEC completing the $ F= 2 \rightarrow  F'= 1  \rightarrow F=2 $ second order virtual transition. (b) A quantum network of cavities connected in the configuration shown in (a).  }
\end{center}
\end{figure}

\section{Entanglement generation}

\label{sec:entgen}

In this section we describe the protocol for entanglement 
generation, where an effective $ S^z_i S^z_j $ interaction is created between two particular nodes $ i,j $ on a quantum network. 
Each of the BECs are placed in a cavity and are connected by optical fiber~\cite{pellizzari97}. Figure \ref{fig1} shows the basic idea of the scheme. BECs are placed in an optical cavity off-resonantly tuned to the  $ F'= 1 \leftrightarrow  F= 2 $ transition, and are mutually connected by optical fiber. Such optical cavities have been already achieved on atom chip systems \cite{colombe07}. 
The BECs on each of the nodes are initially in an arbitrary state of the form (\ref{becqubit}), where the operators  $ a $ and $ b $ are associated with the hyperfine ground states  $ | F= 1, m_F=-1 \rangle $ and  $ | F= 2, m_F=1 \rangle $ respectively. Without the application of lasers on the BECs, the state of the system is stable, as the cavities are off-resonant to the transitions to the excited states. The two qubit interaction can be turned on and off  between two particular nodes as desired by the application of an off-resonant laser tuned between the $  F'= 1  \leftrightarrow  F= 2  $ states.  This completes the second order virtual transition of an atom being excited into the $ F'=1 $ and relaxing back into the original $ F=2 $ state, which allows for the two BEC qubit interaction to be implemented. Since the nodes are only ``on'' when the lasers completing the virtual transition between the $  F'= 1  \leftrightarrow  F= 2  $ states, many such cavities can be connected together into a quantum network (Figure \ref{fig1}b), where entanglement can be generated between nodes on demand. 

A more mathematical description of the scheme is as follows. The cavity is modeled by a single quantized mode coupled to the transition
$ |F'=1,m_F=0\rangle \leftrightarrow |F=2,m_F=+1\rangle $ with coupling strength $G$.  Atoms in the excited state $ |F'=1,m_F=0\rangle  $ are 
denoted $e_i $, and those in state $ |F=2,m_F=+1\rangle$ are denoted $b_i $, where $ i=1,2 $ label the nodes where the entanglement is to 
be generated. The state $ |F=1,m_F=-1\rangle$, labeled as $a_i $, constituting the remaining logical state encoding the BEC qubit state (\ref{becqubit}). The cavity QED Hamiltonian for the bosonic case is
\begin{align}
\label{hamiltonian0}
H_{\mbox{\tiny CQED}} = & \sum_{i=1,2} G(e_i^\dagger b_i p_i+p_i^\dagger b_i^\dagger e_i)+ \hbar \omega_0 e_i^\dagger e_i  +  \hbar \omega p_i^\dagger p_i,
\end{align}
where $ p_i $ are the cavity photon annihilation operators for each node $ i $. The detuning between the transition to the excited state and the cavity photon resonance is $ \Delta = \hbar \omega_0 - \hbar \omega $.  The Hamiltonian describing the 
coupling between the cavity modes and fiber mode reads~\cite{serafini06}
\begin{align}
\label{ham_fib}
H_{\mbox{\tiny f}} = \nu p(p_1^\dagger+e^{i\phi}p_2^\dagger)+\mbox{H.c.},
\end{align}
where $\nu$ is the cavity--fiber coupling strength and $\phi$ is the phase due to propagation of the field through the fiber. We consider the short fiber limit $l\bar{\nu}/2\pi c \leq 1$ where $l$ is the length of the fiber and $\bar{\nu}$ is the decay rate of the cavity field into the continuum of the fiber modes. In this case only the resonant mode $p$ of the fiber interacts with the cavity modes~\cite{serafini06}.  
Using the canonical transformations $c=1/\sqrt2(p_1-e^{-i\phi}p_2),$ $c_1=1/2(p_1+e^{-i\phi}p_2+\sqrt2 p),$ $c_2=1/2(p_1+e^{-i\phi}p_2-\sqrt2 p)$ we can rewrite $H_{\mbox{\tiny f}}$ in diagonal representation
\begin{align}
\label{ham_fib_diag}
H_{\mbox{\tiny f}} = \sqrt2\nu (c_1^\dagger c_1-c_2^\dagger c_2).
\end{align}

Assuming the modes $c_1$ and $c_2$ are off resonant we may then only consider the collective mode $ c $ alone.  The interaction Hamiltonian of whole system is then
\begin{align}
\label{hamiltonian_int}
H_{\mbox{\tiny int}} =H_{\mbox{\tiny CQED}}+H_{\mbox{\tiny f}}= \frac{G}{\sqrt{2}}(e_1^\dagger b_1 c-e^{i\phi}e_2^\dagger b_2 c+H.c.)+ \hbar \omega_0 (e_1^\dagger e_1+e_2^\dagger e_2)  +  \hbar \omega c^\dagger c,
\end{align}
Using adiabatic condition with large detuning $ \Delta \gg G $, we may neglect the effects of very rapidly varying terms and adiabatically eliminate the photons and excited levels from our scheme~\cite{james00,brion07}. Following the standard procedure for adiabatic elimination and setting
$ \frac{d ce_1^\dagger }{dt} = \frac{d ce_2^\dagger }{dt} = 0 $, we obtain
\begin{align}
\label{oscillating_term}
ce_1^\dagger& =-\frac{G}{\sqrt{2}\Delta} \left[ e^{-i\phi}e_1^\dagger e_2 b_2^\dagger-(e_1^\dagger e_1-c^\dagger c)b_1^\dagger\ \right], \nonumber \\
ce_2^\dagger& =-\frac{G}{\sqrt{2}\Delta} \left[ (e_2^\dagger e_2-c^\dagger c)e^{-i\phi}b_2^\dagger-e_1 e_2^\dagger b_1^\dagger \right] .
\end{align}
Substituting the expressions of \eq{oscillating_term} into the expression of \eq{hamiltonian_int} the effective Hamiltonian can be written as
\begin{align}
\label{ham_eff1}
H_{{\mbox{\tiny eff}}} = & -\frac{G^2}{\Delta} \left[ e^{-i\phi}e_1^\dagger e_2 b_2^\dagger b_1+e^{i\phi}b_1^\dagger b_2 e_2^\dagger e_1 \right] 
+\frac{G^2}{\Delta} \left[ e_1^\dagger e_1 b_1^\dagger b_1+ e_2^\dagger e_2 b_2^\dagger b_2\right] 
+\hbar \omega_0 (e_1^\dagger e_1+e_2^\dagger e_2) \nonumber \\
& + 
\left[ \hbar \omega - \frac{G^2}{\Delta} (b^\dagger_1 b_1 + b^\dagger_2 b_2 ) \right] c^\dagger c,
\end{align}
Now considering the effect of the bright pump laser 
\begin{align}
\label{hamiltonian_pump}
H_{\mbox{\tiny pump}}=\sum_{i=1,2} g(e_i b_i^\dagger+b_i e_i^\dagger)+ \Delta e_i^\dagger e_i
\end{align}
where we have assumed that the laser is detuned by the same amount $ \Delta = \hbar \omega_0 - \hbar \omega $.  We may then adiabatically eliminate the excited state by assuming $ \frac{d e_{1,2}}{dt}=0 $, which gives the relations $e_1= -\frac{g}{\Delta }b_1$ and $e_2=-\frac{g}{\Delta }b_2$.  Rewriting this in terms of spin operators $ S^z_i = a_i^\dagger a_i - b_i^\dagger b_i $ and number operators $ {\cal N}_i = a_i^\dagger a_i + b^\dagger_i b_i $ gives up to constant terms
\begin{align}
H_{{\mbox{\tiny eff}}} = & - \hbar \Omega \cos\phi S_{1}^z S_{2}^z + \frac{\hbar \Omega}{2} \left[ (S_{1}^z)^2 + (S_{2}^z)^2 \right] \nonumber \\
& + \left[ \hbar \Omega ( N(\cos \phi -1) + 2\cos \phi) - \frac{g^2 \hbar \omega_0 }{2 \Delta} \right] ( S^z_1 + S^z_2)
\label{effectiveham1}
\end{align}
where 
\begin{align}
\hbar \Omega = \frac{G^2 g^2}{2\Delta^3}  .
\label{effective2qubit}
\end{align}
and have assumed a constant number of bosons $ {\cal N} = N $ for each BEC and the photon number is small. 
We thus see that two qubit interactions can be produced, as well as effective self-interaction terms $ (S^z_i)^2 $ and rotation terms $ S^z_i $.  The single qubit rotation terms are relatively harmless as these may be compensated for assuming single qubit control is available, using existing methods such as that described in Ref. \cite{bohi09}.  However, the self-interaction terms are an unwanted by-product of the procedure and require elimination in order to obtain a pure $ S^z_1 S^z_2 $ interaction.  

To eliminate the unwanted terms, we note that a similar procedure but blocking the optical fiber connection produces the same quadratic terms as (\ref{effectiveham1}) but without the interaction term $ S^z_1 S^z_2 $.  Performing the same adiabatic elimination steps directly on (\ref{hamiltonian0}) by first eliminating $ p_i c_i^\dagger $, then $ e_i $ yields
\begin{align}
H_{{\mbox{\tiny eff}}}' = &  \frac{\hbar \Omega'}{2} \left[ (S_{1}^z)^2 + (S_{2}^z)^2 \right] - \left[ N\hbar \Omega' + \frac{g^2 \hbar \omega_0 }{2 \Delta' } \right] ( S^z_1 + S^z_2).
\label{effectiveham2}
\end{align}
with
\begin{align}
\hbar \Omega' = \frac{G^2 g^2}{{\Delta'}^3}  .
\label{effective2qubit2}
\end{align}
where we have assumed a different value of the detuning for generality. To make the undesired terms cancel we may choose the detuning $ \Delta' = - \Delta $, which reverses the sign of (\ref{effective2qubit2}).  
 The procedure for implementing a pure $ S^z_1 S^z_2 $ interaction is then as follows: (i) Apply (\ref{effectiveham1}) for a desired time $ t= \tau $.  (ii) Apply (\ref{effectiveham2}) with the reverse detuning $ \Delta ' = - \Delta $ for a time $ t= \tau/2 $.  This then implements the total Hamiltonian
\begin{align}
H_{{\mbox{\tiny eff}}}^{\mbox{\tiny tot}} = & -\hbar \Omega \cos\phi S_{1}^z S_{2}^z + \left[ (N+2)\hbar \Omega \cos \phi - \frac{g^2 \hbar \omega_0 }{4 \Delta} \right] ( S^z_1 + S^z_2)
\label{effectivehamtot}
\end{align}
for a time $ \tau $ as desired. An alternative procedure would be to adjust the phase $ \phi $ in (\ref{effectiveham1}) to turn off the $ S^z_1 S^z_2 $ interaction term by choosing $ \phi = \pi/2 $.  By again reversing the detuning allows to cancel the undesired self-interaction terms.

The form of the entanglement is a continuous version of that produced in standard qubits, and together with 
one qubit gates form a universal gate set that can be used to construct quantum algorithms \cite{byrnes12} and 
is required for teleportation schemes \cite{pyrkov13}. The state that is created using the $ S^z_1 S^z_2 $ interaction
is more complicated than for standard qubits, thus is a non-trivial problem to analyze its properties, and the effects of
decoherence(see Ref. \cite{byrnes13} for more details). To illustrate the complexity of the state, consider two BEC qubits in maximal $ x $-eigenstates.  The evolved state is
\begin{equation}
e^{-i \Omega S^z_1 S^z_2 t} | \frac{1}{\sqrt{2}}, \frac{1}{\sqrt{2}} \rangle \rangle_1 
| \frac{1}{\sqrt{2}}, \frac{1}{\sqrt{2}} \rangle \rangle_2 \nonumber = \frac{1}{\sqrt{2^N}} \sum_k  \sqrt{N \choose k} | \frac{e^{i(N-2k) \Omega t}}{\sqrt{2}} , \frac{e^{-i(N-2k) \Omega t}}{\sqrt{2}} \rangle \rangle_1 | k \rangle_2 ,
\label{entangledstate}
\end{equation}
where $ |k \rangle_i = \frac{(a^\dagger_i)^k (b^\dagger_i)^{N-k}}{\sqrt{k! (N-k)!}}| 0 \rangle $ are eigenstates of $ S^z_i $ (Fock states).  
Despite the state's complexity, it was shown in Ref. \cite{byrnes12} that such a Hamiltonian can be used to make the analogue of the 
CNOT operation for BEC qubits. In Ref. \cite{pyrkov13} it was shown that teleportation of spin coherent states could be performed with 
$ \Omega t = 1/\sqrt{2N} $. In this paper we do not study the types of states that are generated, and refer the reader to Refs. \cite{byrnes12,byrnes13}. In section \ref{sec:dec} we will be concerned with how when the scheme described above can reliably create entangled states with various decohering effects.

\section{Decoherence}
\label{sec:dec}

Decoherence effects in BECs and cavity QED model have been the subject of many works~\cite{pellizzari95,burt97,sinatra98,anglin97,ferrini10,ferrini11,trimborn11}. Here we discuss the effect of decoherence on the scheme introduced in the previous section.  Specifically, as excited states such as $ F'=1 $ are used in our proposed scheme, it is important to analyze effects of spontaneous emission, which contributes to decoherence.  Another potential source of decoherence is photon loss from the cavity.  
The way we will approach the problem is to estimate the effects of these processes separately by breaking the problem into parts and examining each of the processes in a prototypical configuration.   This approach will make clear what the important sources of decoherence to the scheme are, and how to avoid them.  Special emphasis will be made on the scaling properties of the decoherence with the atom number $ N $, which is typically a large number in our case. This will identify what kind of parameters need to be chosen for the current scheme.  From this we obtain simple formulas for the decoherence rates for each process. The typical experimentally achievable parameters for BEC entanglement are then estimated using these formulas in section \ref{sec:numbers}.

\subsection{Spontaneous emission}
\label{sec:spon}

\begin{figure}
\begin{center}
\scalebox{0.6}{\includegraphics{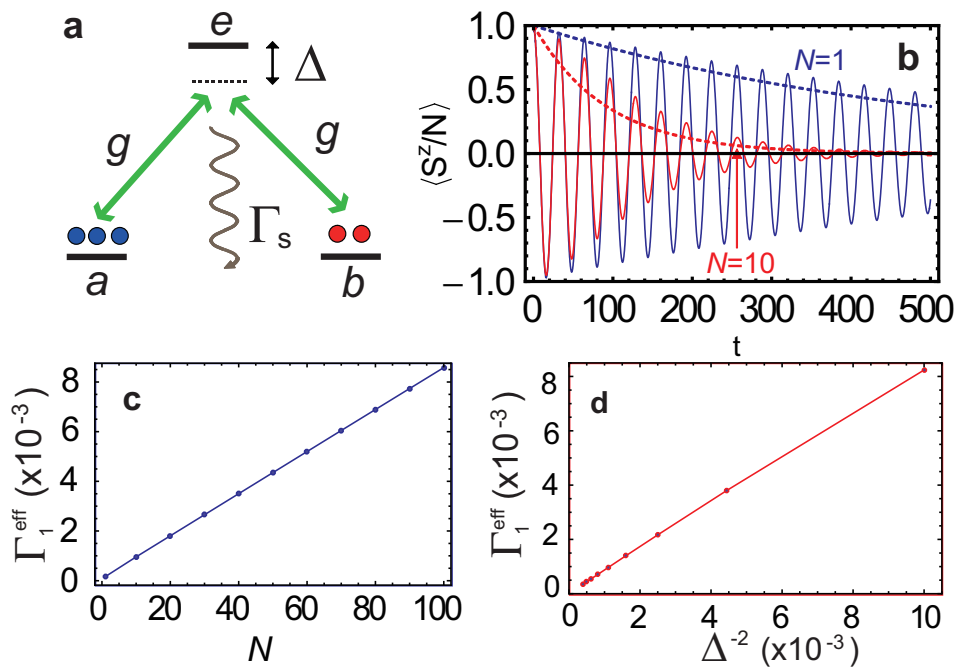}}
\caption{\label{fig3} Decoherence effects due to spontaneous emission.  (a) Schematic for modeling effects of spontaneous emission.  A BEC qubit stored in levels $ a $ and $ b $ is manipulated via an adiabatic passage through an excited state $ e $.  The excited state is susceptible towards spontaneous emission.  (b) Rabi oscillations of the BEC qubit, as measured by the expectation value of $ S^z = a^\dagger a - b^\dagger b $ for $ g=1 $, $ \Delta = 10 $, $ \Gamma_s = 0.1 $ in (\ref{spontaneousmaster}).  The effective decay rate $ \exp[-\Gamma_1^{\mbox{\tiny eff}} t] $ is plotted for comparison (dotted lines). (c) Numerically determined effective decoherence rates $ \Gamma_1^{\mbox{\tiny eff}} $ as a function of boson number $ N $ for $ g=1 $, $ \Delta = 10 $, $ \Gamma_s = 0.01 $.  (d) Numerically determined effective decoherence rates $ \Gamma_1^{\mbox{\tiny eff}} $ as a function of detuning $ 1/\Delta^{2} $ for $ g=1 $, $ N = 100 $, $ \Gamma_s = 0.01 $.  }
\end{center}
\end{figure}

To analyze the effect of spontaneous emission on the coherence of BECs, we consider a minimal model where 
single BEC qubit transitions via an adiabatic passage through an excited state.   The excited state is 
considered to be unstable towards spontaneous emission, where it can decay to either one of the ground states (see Figure \ref{fig3}a). To model this, we consider the master equation
\begin{align}
\frac{d \rho}{dt} = & \frac{i}{\hbar} [ \rho,H_1] - \frac{\Gamma_s}{2} \left[ e^\dagger a a^\dagger e \rho - 2 a^\dagger e \rho e^\dagger a +
\rho e^\dagger a a^\dagger e \right] \nonumber \\
& - \frac{\Gamma_s}{2} \left[ e^\dagger b b^\dagger e \rho - 2 b^\dagger e \rho e^\dagger b +
\rho e^\dagger b b^\dagger e \right]
\label{spontaneousmaster}
\end{align}
where $ H_1 $ is 
\begin{align}
H_1 = & \Delta e^\dagger e  + g ( a^\dagger e + e^\dagger a) +  g (b^\dagger e + e^\dagger b)  .
\label{singlequbitadiabatic}
\end{align}
Here we have assumed for simplicity that the coupling $ g $ to the intermediate level from the logical states $ a $ and $ b $ are equal, and $ \Delta $ is the detuning of the laser between states $ a,b$ to the excited state $ e $.  To a very good approximation, Eq. (\ref{spontaneousmaster}) can be solved for arbitrary $ N $. From the Heisenberg equations of motion, we may derive equations for the expectation values
\begin{align}
\frac{\langle a^\dagger a \rangle}{dt} & = -i\frac{g}{\hbar}(\langle a^\dagger e \rangle-\langle e^\dagger a \rangle) + \Gamma_s \langle e^\dagger e \rangle (\langle a^\dagger a \rangle +1 ) \nonumber \\
\frac{\langle a^\dagger e \rangle}{dt} & = -i\frac{g}{\hbar}(\langle a^\dagger a \rangle -\langle e^\dagger e \rangle +\langle a^\dagger b \rangle ) -i \frac{\Delta}{\hbar} \langle a^\dagger e \rangle \nonumber \\
& - \frac{\Gamma_s}{2}( \langle a^\dagger a \rangle -\langle e^\dagger e \rangle+\langle b^\dagger b \rangle +1)\langle a^\dagger e\rangle \nonumber \\
\frac{\langle e^\dagger e \rangle}{dt} & = -i\frac{g}{\hbar}(\langle e^\dagger a \rangle-\langle a^\dagger e \rangle -\langle b^\dagger e \rangle + \langle e^\dagger b \rangle ) \nonumber \\
& -\Gamma_s (\langle a^\dagger a \rangle+1)\langle e^\dagger e \rangle-\Gamma_s (\langle b^\dagger b\rangle+1)\langle e^\dagger e \rangle \nonumber \\
\frac{\langle a^\dagger b \rangle}{dt} & = i\frac{g}{\hbar}(\langle e^\dagger b \rangle - \langle a^\dagger e \rangle ) 
+ \Gamma_s \langle e^\dagger e \rangle \langle a^\dagger b \rangle \nonumber \\
\frac{\langle b^\dagger e \rangle}{dt} & = -i\frac{g}{\hbar}(\langle b^\dagger a \rangle + \langle b^\dagger b \rangle -\langle e^\dagger e \rangle) - i  \frac{\Delta}{\hbar} \langle b^\dagger e \rangle \nonumber \\
& - \frac{\Gamma_s}{2} (\langle a^\dagger a \rangle + 1) \langle b^\dagger e \rangle
- \frac{\Gamma_s}{2} (\langle b^\dagger b \rangle - \langle e^\dagger e \rangle) \langle b^\dagger e \rangle \nonumber \\
\frac{\langle b^\dagger b \rangle}{dt} & = i\frac{g}{\hbar}( \langle e^\dagger b \rangle - \langle b^\dagger e \rangle ) +  \Gamma_s \langle e^\dagger e \rangle (\langle b^\dagger b \rangle +1) 
\label{spontaneouseqs}
\end{align}
where quartic products of bosonic operators were approximated by products of quadratic correlations.  This allows for a closed set of equations (\ref{spontaneouseqs}), which together with their complex conjugates can be solved numerically.  Comparison of the exact solutions of (\ref{spontaneousmaster}) for small boson numbers show virtually perfect agreement with (\ref{spontaneouseqs}).

Results of the simulation are shown in Figure \ref{fig3}b. We see that for large detunings $ \Delta \gg g $, Rabi oscillations occur between the levels $ a $ and $ b $ with a characteristic frequency 
\begin{align}
\hbar \Omega_1^{\mbox{\tiny eff}} = \frac{g^2}{\Delta} .
\label{sponeffg}
\end{align}
The spontaneous emission introduces a decoherence to the oscillations, which increase with $ N $.  From the decay envelope of the oscillations, it is possible to estimate the decoherence rates induced by the spontaneous emission.  We find almost perfect linear behavior in $ N $ and inverse quadratic behavior in the detuning $ \Delta $  (Figures \ref{fig3}c and \ref{fig3}d).  The rate of the decoherence can be simply summarized by the formula
\begin{align}
\label{effectiveonequbitdecoherence}
\Gamma_1^{\mbox{\tiny eff}} \approx \frac{g^2 \Gamma_s (N+1)}{\Delta^2} .
\end{align}
The  decay envelope $ \exp[-\Gamma_1^{\mbox{\tiny eff}} t ] $ is plotted together with the simulated results in Figure \ref{fig3}b.  We see that the (\ref{effectiveonequbitdecoherence}) gives good quantitative agreement with the data. 

To understand the origin of the proportionality with $ N+1 $ of the decoherence rate, let us consider a simpler toy model of
spontaneous decay between two levels populated by $ N $ atoms.  In this case the master equation is 
\begin{align}
\frac{d \rho}{dt} = &- \frac{\Gamma_s}{2} \left[ e^\dagger a a^\dagger e \rho - 2 a^\dagger e \rho e^\dagger a +
\rho e^\dagger a a^\dagger e \right] 
\label{simplespontaneousmaster}
\end{align}
where $ e $ is an excited state and $ a $ is a ground state. Noting that the master equation conserves total particle number $ N = a^\dagger a + e^\dagger e $, it can be solved by finding the equations of motion
\begin{align}
\frac{d \langle e^\dagger e \rangle }{dt} = - \Gamma_s (N+1) \langle e^\dagger e \rangle + \Gamma_s \langle e^\dagger e \rangle^2
\end{align}
where we have used the approximation $ \langle (e^\dagger e)^2 \rangle \approx \langle e^\dagger e \rangle^2 $.  
This allows for a solution
\begin{align}
\langle (e^\dagger e - a^\dagger a)  \rangle = N \tanh \left[ -\frac{\Gamma_s (N+1)t}{2} + K_0 \right] ,
\label{spon2qubit}
\end{align}
where $ K_0 $ is a constant depending upon the initial conditions.  We see that the 
time constant within the hyperbolic function is proportional to $ N +1 $, which can be attributed to bosonic final state stimulation \cite{byrnes11}.  The effect of a large number of bosons is that it enhances the dissipation rate $ \Gamma_s \rightarrow \Gamma_s (N+1) $.  For the lambda scheme (\ref{spontaneousmaster}) examined above this results in a decoherence rate of the Rabi oscillations also accelerated by a factor $ N+1 $.

The linear scaling of (\ref{effectiveonequbitdecoherence}) with $ N $ may seem detrimental to the effectiveness of the two qubit gate proposed in the previous section, as $ N $ is typically a very large number in typical BECs.  However, the quadratic scaling of $ \Gamma_1^{\mbox{\tiny eff}}$ with $ \Delta $ allows for a compensation of the large $ N $ via increasing the detuning to a sufficient amount.  This naturally decreases the Rabi frequency $ \Omega_1^{\mbox{\tiny eff}} $, so one must choose the parameters to optimize this trade-off.  Numerical parameters which satisfy this will be shown in section \ref{sec:numbers}.

\subsection{Cavity photon loss}

\begin{figure}
\begin{center}
\scalebox{0.6}{\includegraphics{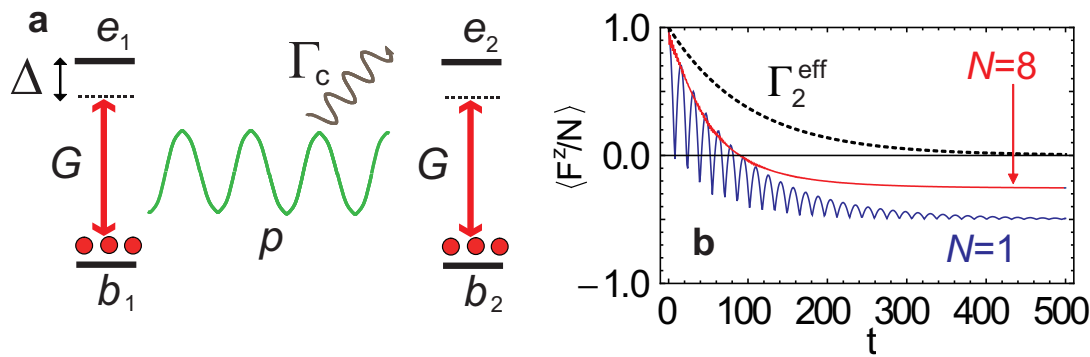}}
\caption{\label{fig4} (a) Schematic for a two BEC qubit entangling gate operation with photon loss. The BEC qubit is stored between the states $ a_i $ and $ b_i $ ($i=1,2$).  The entangling operation occurs via a photon being emitted from one of the BEC qubits, say node $ i = 1 $.  The photon is absorbed by the other node $ i=2 $.  (b) Decoherence induced by cavity photon loss for repeated applications and reversals of the two qubit operation for gate times $ \Omega_2^{\mbox{\tiny eff}} t = \frac{\pi}{4N} $. Parameters used are $ \Gamma_c = 1 $, $ G= 1 $, $ \Delta = 10 $ in (\ref{cavityphotondecay}). The dotted line shows the effective decay rate $ \exp[-\Gamma_2^{\mbox{\tiny eff}} t] $.}
\end{center}
\end{figure}

Another mechanism of decoherence is via photon loss from the cavity and optical fiber. The model that we will consider is shown in Figure \ref{fig4}a with the master equation
\begin{align}
\label{cavityphotondecay}
\frac{d \rho}{dt} = \frac{i}{\hbar} [ \rho,H_{2}] - \frac{\Gamma_c}{2} \left[ p^\dagger p \rho - 2 p \rho p^\dagger +
\rho p^\dagger p \right]
\end{align}
where 
\begin{align}
\label{qubushamiltonian}
H_2 = \hbar \omega_0 \sum_{n=1,2} e_n^\dagger e_n + \hbar \omega p^\dagger p + G  \sum_{n=1,2} \left[ F^-_n p^\dagger + F^+_n p \right] .
\end{align}
where $ F^z_n = e_n^\dagger e_n - b_n^\dagger b_n$, $ F^+_n = e_n^\dagger b_n  $,  $ \omega_0 $ is the transition frequency, and $ p $ is the photon annihilation operator. The basis of considering such a model is as follows. Starting from (\ref{hamiltonian0}) we may derive (\ref{cavityphotondecay}) by substituting the diagonalized expressions for the photon mode in (\ref{ham_fib}).  
In the protocol of section \ref{sec:entgen}, there are additional lasers corresponding to the $ g $ transitions in Figure \ref{fig1}a.  The purposes of these are twofold. Firstly they are to avoid storage of the quantum information in excited states, which give undesirable effects of spontaneous emission as discussed in the previous section.  Due to the adiabatic passage through the excited state, the population in these states are kept to a very low level at all times, which mitigates the spontaneous emission problem. Secondly, the lasers give on/off control to the two BEC qubit gate such that entanglement can be generated between nodes as desired. In our current analysis we may ignore both of these aspects as we would like to concentrate only on the effects of photon loss to the fidelity of the two qubit gate.  Thus we may consider the effect of the laser to connect levels $ a_n $ and $ e_n $, giving rise to the Hamiltonian (\ref{qubushamiltonian}).  

The basic effect of (\ref{qubushamiltonian}) may be understood by examining the limit of very large detuning $ \Delta = \hbar \omega_0 - \hbar \omega $.  By adiabatically eliminating the photons in a similar way to Ref. \cite{byrnes12}, we obtain the effective Hamiltonian
\begin{align}
H^{\mbox{\tiny eff}}_{2} = \hbar \Omega_2^{\mbox{\tiny eff}} (F^+_1 + F^+_1)(F^-_2 + F^-_2) + \mbox{H.c.},
\label{simpleentangler}
\end{align}
where 
\begin{align}
\hbar \Omega_2^{\mbox{\tiny eff}} = \frac{G^2}{\Delta} .
\label{caveffg}
\end{align}
Eq. (\ref{simpleentangler}) is an entangling two BEC qubit Hamiltonian.  We expect that the cavity photon decay will add a decohering effect to the dominant effect described by (\ref{simpleentangler}).  

We solve (\ref{cavityphotondecay}) by time evolving the density matrix numerically.  Assuming $ N $ bosons per node $ n =1,2 $, the basis states
\begin{align}
| k_1, k_2, n \rangle = \frac{1}{\sqrt{k_1 ! (N-k_1)! k_2! (N-k_2)! n! }} \nonumber
(b_1^\dagger)^{k_1} (e_1^\dagger)^{N-k_1} (b_2^\dagger)^{k_2} (e_2^\dagger)^{N-k_2} (p^\dagger)^n | 0 \rangle
\end{align}
may be taken. The number of basis states may be restricted by noticing that the $ H_2 $ conserves the number of excited states
\begin{align}
n_{\mbox{\tiny ex}} = 2N - k_1 -k_2 +n .
\end{align}
This allows us to eliminate $ k_2 $ in favor of $ n_{\mbox{\tiny ex}} $ and use an alternative basis set
\begin{align}
| k, n, n_{\mbox{\tiny ex}}  \rangle' \equiv | k_1=k, k_2=2N - k_1 -n_{\mbox{\tiny ex}} +n, n \rangle.
\end{align}
Then given an excitation sector $ n_{\mbox{\tiny ex}} $ we may deduce that the only allowed states are $ \max (0,N- n_{\mbox{\tiny ex}}) \le k \le N $ and $0 \le n \le n_{\mbox{\tiny ex}} - N + k$.  For very large detunings $ \Delta = \hbar \omega_0 -\hbar \omega $, we may further assume that the photon number is very small and safely assume a cutoff on $ n $. In our numerical calculations the cutoff is assumed to be $ n = 1$.  

The interaction (\ref{simpleentangler}) in general produces entanglement between the two nodes, which is in general a complicated state for BEC qubits as seen in (\ref{entangledstate}).  It is therefore not obvious how to evaluate the fidelity of the operation. To give a quantitative evaluation of the effects of decoherence due to cavity decay we therefore perform the following procedure: (i) Initialize the state to $ | k_1=0, k_2=N, n=0 \rangle $; (ii) Evolve the $ H_2 $ for a time $ \tau $, which generates an entangled state between the two nodes; (iii) Reverse the Hamiltonian (i.e. evolve $ -H_2 $ for the same time $ \tau $, with the same cavity photon decay; (iv) Repeat steps 2 and 3 repeatedly and evaluate $ \langle F^z_n \rangle $ after each evolution. 
Under ideal conditions with no photon decay, the successive application of $ H_2 $ and $ -H_2 $ first creates an entangled state, then returns this to the initial state.  However, with photon decay this process cannot perfectly, and this will appear as a reduced fidelity of $ \langle F^z_n \rangle $ after the evolution.

In Figure \ref{fig4}b we show $ \langle F^z_1 \rangle $ for the above procedure.  As expected the curves decay in amplitude as expected due to decoherence induced by cavity photon decay. Using similar methods as in section \ref{sec:spon}, we find that the curves have an effective decoherence rate of 
\begin{align}
\Gamma_2^{\mbox{\tiny eff}} \approx \frac{G^2 \Gamma_c}{\Delta^2} .
\label{cavityphotondecoherence}
\end{align}
This expression has a similar form to (\ref{effectiveonequbitdecoherence}) with the exception that it is independent of $ N $. Why is (\ref{cavityphotondecoherence}) independent of $ N $, while (\ref{effectiveonequbitdecoherence}) had an factor accelerating the decoherence by $ N +1 $? To understand this let us consider a single isolated cavity the population of photons $ n = p^\dagger p  $ under the master equation 
\begin{align}
\label{simplecavityphotondecay}
\frac{d \rho}{dt} = - \frac{\Gamma_c}{2} \left[ p^\dagger p \rho - 2 p \rho p^\dagger +
\rho p^\dagger p \right]
\end{align}
obeys 
\begin{align}
\langle p^\dagger p \rangle = n_0 e^{-\Gamma_c t} , 
\end{align}
where $ n_0 $ is the initial photon population.  
We note that the exponent is independent of the initial population.  Physically, the reason for this is that when a photon escapes a cavity, it exits into the vacuum state of the electromagnetic field, which has by definition zero photons. For spontaneous emission there is always $ \sim O(N) $ bosons in the final state, thus transitions are also accelerated by this factor.  This explains why there is no  bosonic final state stimulation in this case, and thus no acceleration of (\ref{cavityphotondecoherence}) due to the boson number $ N $.

\subsection{Effects of dephasing during two BEC qubit operation}

In addition to the effects of spontaneous emission and photon decay, the BEC qubits will experience dephasing which may originate
from many sources, such as experimental fluctuations in the traps for the BECs. Typical dephasing times have been estimated to be of the order of seconds \cite{treutlein06}, hence we expect that this will be less significant in comparison to the effects described above.  However, due to the 
complex state structure of the two BEC entangling operation, we will find that there is a strong dependence of the effects of decoherence upon the gate times of the entangling operation.  This will show that certain states produced in (\ref{entangledstate}) will be more difficult to produce than others. 

\begin{figure}
\begin{center}
\scalebox{0.6}{\includegraphics{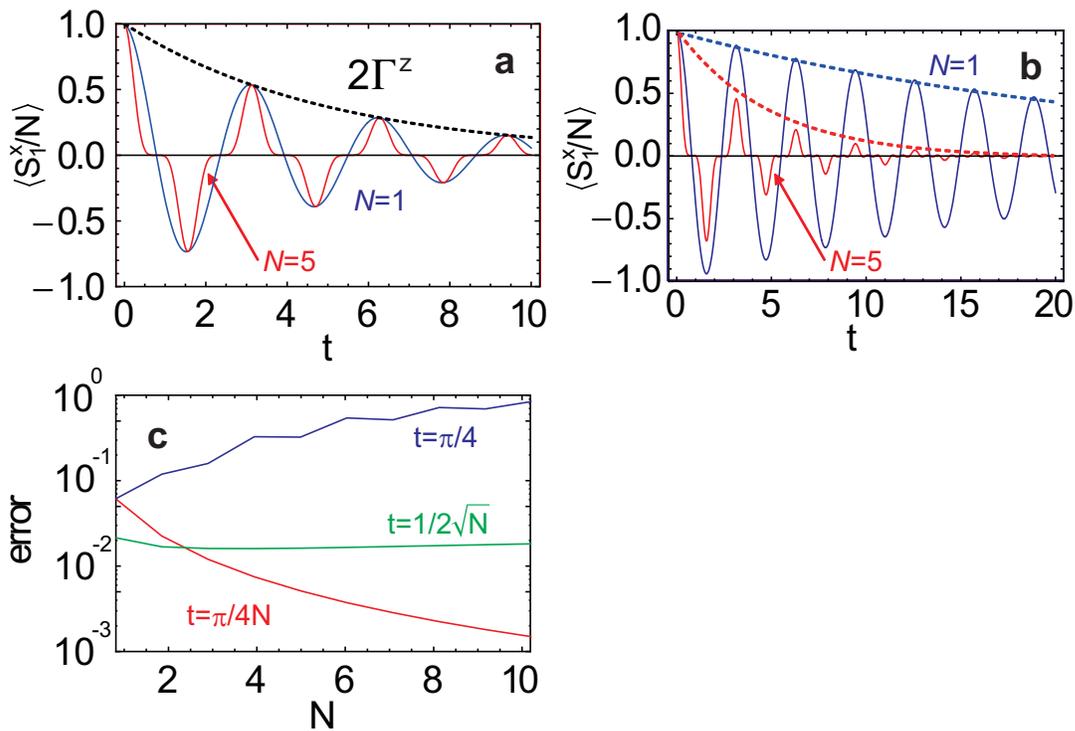}}
\caption{\label{fig5} Decoherence effects of using bosonic qubits for coherent operations. (a) Evolution under a two qubit  $ S^z_1 S^z_2 $ gate under $ S^z $-type decoherence for $ \Gamma_z = 0.1 $.   (b) Evolution under a two qubit  $ S^z_1 S^z_2 $ gate under $ S^x $-type decoherence for $ \Gamma_z = 0.01 $. The effective decoherence envelope $ \exp[-\Gamma^{\mbox{\tiny eff}}_x t] $ is shown as the dotted lines.  (c) The error of the two qubit operation as described in as function of boson number $ N $ for various evolution times as shown.  All calculations assume use $ \Omega = 1 $.  }
\end{center}
\end{figure}

To illustrate this effect, we consider the dephasing master equation
\begin{align}
\label{twoqubitdephasing}
\frac{d \rho}{dt} = i \Omega  [ \rho, S^z_1 S^z_2 ] - \frac{\Gamma_j}{2} \sum_{n=1}^2 [ (S_n^j)^2 \rho - 2 S_n^j \rho S_n^j + \rho (S_n^j)^2 ] ,
\end{align}
where $ j = x,z $.  Considering dephasing in the $ j=z $ direction first, it is possible to analytically solve (\ref{twoqubitdephasing}) given an initial state $ \rho_0 $, yielding
\begin{align}
\rho(t) = & \sum_{k_1,k_2,k_1',k_2'} \rho_0(k_1,k_2,k_1',k_2') e^{-2 \Gamma_z ( (k_1-k_1')^2 + (k_2-k_2')^2)t}  \nonumber \\
& \times e^{-i(2k_1 -N)(2 k_2 -N) \Omega t+i(2k_1' -N)(2 k_2' -N)  \Omega t} |k_1, k_2 \rangle \langle k_1', k_2'| .
\end{align}
Initially preparing the state $ | \frac{1}{\sqrt{2}}, \frac{1}{\sqrt{2}} \rangle \rangle_1 | \frac{1}{\sqrt{2}}, \frac{1}{\sqrt{2}} \rangle \rangle_2 $, evolving under (\ref{twoqubitdephasing}), and calculating the expectation values
$ \langle S^x_1 \rangle $ yields Figure \ref{fig5}a.  We see that at points corresponding to integer multiples of $ t=\pi/2 $, the curves for different $ N $ coincide. This is in fact true for all $ N $, and is possible to derive the relation
\begin{align}
\langle S^x_i(t) \rangle_{\Gamma_z} = e^{-2 \Gamma_z t } \langle S^x_i(t) \rangle_{\Gamma_z=0}.
\end{align}
That is, in the presence of $ z $-decoherence, the expectation value of $ S^x $ is simply multiplied by a factor of $ e^{-2 \Gamma_z t }$. One would naively conclude that decoherence does not have an $ N $ dependence, and is not enhanced due to the large number of bosons in the BEC.  However, we shall see below that this is a special case, as we happened to consider a decoherence operator which commutes with the entangling Hamiltonian, which means that the dynamics of the decoherence and interaction are independent.    

For the above reasons it is more instructive to consider the case where the decoherence and the entangling Hamiltonian does not commute, i.e. $ j = x $. In Figure \ref{fig5}b we show results showing the expectation value of $ \langle S_1^x \rangle $ given the same initial conditions as Figure \ref{fig5}a. We see that in this case there is a degradation of the oscillations with increasing $ N $. We phenomenologically find that the effective decoherence rate of the peaks obeys 
\begin{align}
\Gamma^{\mbox{\tiny eff}}_x = 4 N \Gamma_x .
\label{accdec}
\end{align}
Some insight to the origin of this enhanced decoherence can be obtained by examining the states that are produced 
at particular entangling times.  For example, with zero decoherence and at time $ \Omega t =\pi/4 $, the state (\ref{entangledstate}) can be written \cite{byrnes13}
\begin{align}
& e^{-i S^z_1 S^z_2 \frac{\pi}{4}} | \frac{1}{\sqrt{2}}, \frac{1}{\sqrt{2}} \rangle \rangle_1 
| \frac{1}{\sqrt{2}}, \frac{1}{\sqrt{2}} \rangle \rangle_2 = \nonumber \\
& \frac{1}{2} \left( | \frac{1}{\sqrt{2}} , \frac{1}{\sqrt{2}} \rangle \rangle_1  + | \frac{1}{\sqrt{2}} , -\frac{1}{\sqrt{2}} \rangle \rangle_1  \right) | \frac{e^{i \pi N/4}}{\sqrt{2}} , \frac{e^{-i \pi N/4}}{\sqrt{2}} \rangle \rangle_2 \nonumber \\
& + \frac{1}{2} \left( | \frac{1}{\sqrt{2}} , \frac{1}{\sqrt{2}} \rangle \rangle_1  - | \frac{1}{\sqrt{2}} , -\frac{1}{\sqrt{2}} \rangle \rangle_1  \right) | -\frac{e^{i \pi N/4}}{\sqrt{2}} , \frac{e^{-i \pi N/4}}{\sqrt{2}} \rangle \rangle_2 .
\label{schrodingercat}
\end{align}
This is a entangled Schrodinger cat state, and is generally very susceptible to decoherence.  Thus when the states created are of this fragile nature, the effects of decoherence are enhanced due to the large number of bosons, and decay with the enhanced rate (\ref{accdec}).  

Fortunately, for quantum information processing applications, states such as (\ref{schrodingercat}) are typically not necessary. In Ref. \cite{byrnes12} it was found that the analogue of the CNOT operation could be implemented with gate times $ \Omega t = \pi/4N $.  For quantum teleportation, gate times of $\Omega   t =1/\sqrt{2N} $ are required \cite{pyrkov13}. Therefore it is 
of more practical relevance to understand the effects of the decoherence at these gate times, instead of the long gate times of $ t \sim O(1) $ as plotted in Figs. \ref{fig5}a and \ref{fig5}b.  

We may study the effect of the decoherence on these short times by applying the same strategy as that used in cavity photon loss in Figure \ref{fig4}b. Namely we perform the following sequence: (i) 
Evolve (\ref{twoqubitdephasing}) with $ j=x $ for a time $ t $; (ii) Evolve the resulting state under (\ref{twoqubitdephasing}) with $ j=x $, but with $ -S^z_1 S^z_2 $ as the Hamiltonian for the same time $ t $; 
(iii) Measure the error, defined to be $1- \frac{\langle S_1^x \rangle}{N}$.  

The effect of decoherence for various gate times is shown in Figure \ref{fig5}c. We see that the error for various gate times has very different behavior. Specifically, for gate times of order $ \Omega  t = \pi/4 $ the errors increase, as suggested in the results of Fig. \ref{fig5}b.  However, for gate times $ \Omega t = 1/2\sqrt{N} $ the errors stay mostly constant with $ N $, and for short gate times the 
errors decrease with $ N $. This can be understood to be due to the gate time reducing for $ \Omega t = \pi/4N $, which 
results in less time for the dephasing to take effect.  The border between these two behaviors can be seen to be $ \Omega t \sim 1/\sqrt{N} $.   

The above results show that states with $\Omega  t < 1/\sqrt{N} $ are relatively stable against decoherence, while states with $ \Omega t >  1/\sqrt{N} $ are more delicate due to the enhanced decoherence effects originating from the large number of atoms in the BEC.  The origin is due to the nature of the states that are produced.  For gate times $\Omega  t \sim O(1) $,  Schrodinger cat-like states (\ref{schrodingercat}) tend to be produced, which are well-known to be highly susceptible to decoherence. We note that similar results were found in Ref. \cite{byrnes13}.

\section{Estimated gate times}
\label{sec:numbers}

We now estimate some numbers for achievable gate times for the two BEC qubit gate using currently available technology.  Our 
primary constraints are in the quality of the cavities, for which we assume parameters given in Ref. \cite{colombe07}. In our current notation this corresponds to $ G_0/\hbar =  1350 \mbox{ MHz}$, $ \Gamma_s =  19 \mbox{ MHz} $, and $ \Gamma_c =  330 \mbox{ MHz} $, where 
$ G_0 $ is the single atom cavity coupling. Since there are a large number of BEC atoms in the cavity, in our case 
\begin{align}
G = \sqrt{N} G_0
\end{align}
where the $ \sqrt{N} $ comes from the collective atomic enhancement \cite{colombe07}.  In order to effectively perform the adiabatic passage through the excited state in Fig. \ref{fig1}a, it is optimal to choose the control
lasers to have the same strength as the cavity. We thus have a further constraint
\begin{align}
g = G.
\end{align}
To overcome spontaneous emission, we need to choose the detuning sufficiently large such that the factor of $ N + 1 $ in the numerator of (\ref{effectiveonequbitdecoherence}) is canceled off. Choosing $ \Delta = D G_0 N $, and substituting into (\ref{sponeffg}), (\ref{effectiveonequbitdecoherence}), (\ref{caveffg}), and (\ref{cavityphotondecoherence}) we obtain
\begin{align}
\Omega_1^{\mbox{\tiny eff}} & = \frac{G_0}{\hbar D}  = 1350 \mbox{ MHz} \nonumber \\
\Gamma_1^{\mbox{\tiny eff}} & = \frac{\Gamma_s}{D^2} =  19 \mbox{ MHz} \nonumber\\
\Omega_2^{\mbox{\tiny eff}} & = \frac{G_0}{\hbar D}  = 1350 \mbox{ MHz} \nonumber \\
\Gamma_2^{\mbox{\tiny eff}} & = \frac{\Gamma_c}{D^2  N} = 0.3 \mbox{ MHz} 
\label{decoestimates}
\end{align}
where $ D $ is a dimensionless detuning parameter which may be freely chosen to adjust the parameters above. For the numerical estimates above, $ D=1 $ and $ N =1000 $ are chosen.  We see that the coherent coupling greatly exceeds the effective decoherence values, showing that the processes of spontaneous emission and photon decay may be overcome by a suitable choice of detuning.  

For the combined interaction, substituting numbers into (\ref{effective2qubit}) we estimate
\begin{align}
\Omega & = \frac{G_0 }{2\hbar D^3  N} = 0.7 \mbox{MHz}
\end{align}
which may appear on first glance appear to be a poor result in comparison to the decoherence rates in (\ref{decoestimates}).  However, as discussed
in the previous section, for quantum information processing applications, only short gate times are typically necessary.
Such gates can be completed within the decoherence times shown. For example, to perform the analogue of the CNOT gate, interaction times of  $ \Omega t_{\mbox{\tiny CNOT}} = \frac{\pi}{4N} $  are required.  Calculating the ratio of the timescale for decoherence due to spontaneous decay  $ 1/\Gamma_1^{\mbox{\tiny eff}} $ to gate time, we estimate
\begin{align}
\frac{1/\Gamma_1^{\mbox{\tiny eff}}}{t_{\mbox{\tiny CNOT}}} =  44, 
\label{ratioestimate}
\end{align}
showing many CNOT gates can be produced within the decoherence time. 
It is therefore reasonable to expect that our proposed scheme can be used to create entangled states despite the enhanced spontaneous decay rates due to the large number of atoms in the BEC.

\section{Conclusions}
\label{sec:conc}

We have proposed a method for generating $ S^z_i S^z_j $ interaction between two-component BECs using a system of cavities coupled by optical fiber.  The scheme uses a ``quantum bus'' of photons to effectively couple BECs together, and is controllable between any two nodes on the quantum network via local lasers.  The main enemies of the scheme are spontaneous emission and photon loss, which creates an effective decoherence.  In particular, spontaneous decay is enhanced by a factor of  $\sim N $, the number of bosons in the BEC.  Despite this, by choosing a sufficiently large detuning it was shown that two BEC qubit gates can be created within the estimated decoherence times. 
It was also shown that states with $ \Omega t < 1/\sqrt{N} $ should be stable against decoherence while those with $ \Omega t > 1/\sqrt{N} $ tend to be more fragile. This is due to the Schrodinger cat-like states that are produced for longer times, which are more susceptible to decoherence \cite{hecht04}. An alternative scheme would involve not using excited states at all, and using microwave cavities such as that proposed in Refs. \cite{henschel10}. This would reduce spontaneous decay effects and in this case the dominant decoherence mechanism would result from cavity decay.  The formalism developed here could still be equally applied by a simple modification of the definitions assumed here.

\ack

We thank S. Kumar, R. Schmied, S. Koyama for discussions. This work is supported by the Transdisciplinary Research Integration Center, Okawa Foundation, and the Japan Russia Youth Exchange Center and the Center for the Promotion of Integrated Sciences (CPIS) of Sokendai.


\section*{References}
\begin{thebibliography}{99}

\bibitem{kimble08} H.J. Kimble, Nature {\bf 453}, 1023 (2008).

\bibitem{duan10} L.-M. Duan, C. Monroe, Rev. Mod. Phys. {\bf 82}, 1209 (2010).

\bibitem{ladd10} T. D. Ladd,F.  Jelezko, R.  Laflamme, Y. Nakamura, C.  Monroe,J. L. O'Brien,
Nature {\bf 464}, 45 (2010).

\bibitem{eisaman11} M.D. Eisaman {\it et. al.}, Rev. Sci. Instrum. {\bf 82}, 071101 (2011).

\bibitem{duan01} L.-M. Duan {\it et. al.}, Nature {\bf 414} 413 (2001).

\bibitem{ritter12} S. Ritter {\it et.al.}, Nature {\bf 484} 195 (2012).


\bibitem{bohi09} P. B{\"o}hi {\it et al.}, Nature Phys. {\bf 5}, 592 (2009).

\bibitem{riedel10} M. Riedel {\it et al.}, Nature {\bf 464}, 1170 (2010). 

\bibitem{steck08}  D. A Steck, {\it Rubidium 87 D line data}  http://steck.us/alkalidata (Version 2.1.4, last revised 23 December 2010).

\bibitem{colombe07} Y. Colombe {\it et al.}, Nature {\bf 450}, 272 (2007). 

\bibitem{lettner11} M. Lettner {\it et.al.}, Phys. Rev. Lett. {\bf 106} 210503 (2011).


\bibitem{byrnes12} T. Byrnes, K. Wen and Y. Yamamoto, Phys. Rev. A 85, 040306(R) (2012).

\bibitem{gross12} C. Gross {\it J. Phys. B: At. Mol. Opt. Phys.} {\bf 45}, 103001 (2012). 

\bibitem{pyrkov13} A. Pyrkov and T. Byrnes, arxiv: 1305.2479.  

\bibitem{treutlein06b} P. Treutlein, T. W. H{\"a}nsch, J. Reichel, A. Negretti,  M. A. Cirone, T. Calarco, Phys. Rev. A  {\bf 74}, 022312
(2006).

\bibitem{pellizzari97} T. Pellizzari, Phys. Rev. Lett. {\bf 79} 5242 (1997).

\bibitem{serafini06} A. Serafini, S. Mancini and S. Bose, Phys. Rev. Lett. {\bf 96} 010503 (2006).

\bibitem{james00} D.F.V. James, Fortschr. Phys. {\bf 48} 823 (2000).

\bibitem{brion07} E. Brion, L.H. Pedersen and K. Molmer, J. Phys. A: Math. Theor. {\bf 40}, 1033 (2007).

\bibitem{byrnes13} T. Byrnes, arxiv: 1305.5095.





\bibitem{pellizzari95} T. Pellizzari, S. A. Gardiner, J. I. Cirac, and P. Zoller, Phys. Rev.
Lett. {\bf 75}, 3788 (1995).

\bibitem{burt97} E.A. Burt et. al., Phys. Rev. Lett. 79, 337 (1997). 

\bibitem{sinatra98} A. Sinatra and Y. Castin, Eur. Phys. J. D 4, 247 (1998).

\bibitem{anglin97} J. Anglin, Phys. Rev. Lett. 79, 6 (1997).

\bibitem{ferrini10} G. Ferrini et. al., Phys. Rev. A 82, 033621 (2010). 

\bibitem{ferrini11} G. Ferrini et.al., Phys. Rev. A 84, 043628 (2011).

\bibitem{trimborn11} F. Trimborn et. al., Eur. Phys. J. D 63, 63 (2011).

\bibitem{byrnes11}
T. Byrnes, K. Yan, Y. Yamamoto, New J. Phys. {\bf 13}, 113025 (2011). 





\bibitem{treutlein06}  P. Treutlein, T. Steinmetz, Y.  Colombe, B. Lev, P. Hommelhoff, J. Reichel, M. Greiner, O. Mandel, A.  Widera, T. Rom, I. Bloch, T. W.  H{\"a}nsch,  Fortschr. Phys. {\bf 54}, 702 (2006). 



\bibitem{hecht04} T. Hecht, Diploma Thesis, Technische Universit{\"a}t M{\"u}nchen Max-Planck-Institut f{\"u}r Quantenoptik (2004). 

\bibitem{henschel10} K. Henschel, J. Majer, J. Schmiedmayer, H. Ritsch, Phys. Rev. A {\bf 82}, 033810 (2010).  

\end{thebibliography}
\end{document}